\documentstyle[11pt,newpasp,twoside,epsf]{article}
\def\edcomment#1{\iffalse\marginpar{\raggedright\sl#1\/}\else\relax\fi}
\reversemarginpar

\begin{document}
\title{New Circumstellar Magnetic Field Diagnostics}
 \author{K. H. Nordsieck}
\affil{Department of Astronomy, University of Wisconsin - Madison, 475 N. Charter St.,
Madison WI 53706}

\begin{abstract}
In this paper I will discuss new magnetic field diagnostics and instrumentation for an area of
astrophysics where magnetic field observations have been difficult -  circumstellar material. 
Such diagnostics would be particularly relevant to star formation and evolution.  Stellar
photosphere diagnostics include the Zeeman effect and atomic scattering diagnostics like the
Hanle Effect and atomic alignment.  The Zeeman Effect is in general not sensitive enough for the
field strengths expected for circumstellar material, and it is easily defeated by Doppler
broadening in a dynamic envelope.  Atomic scattering diagnostics, pioneered recently for the
Sun, are promising, but have never been applied outside the Sun.  For the study of unresolved
envelopes, the Hanle Effect may be applicable particularly in the ultraviolet.  A medium
resolution UV spectropolarimeter, for instance, would serve for such studies.  Atomic alignment
effects could utilize a ground-based, high signal-to-noise spectropolarimeter, with profile
information from high spectral resolution. I will briefly mention several instrumentation
development efforts in these directions.
\end{abstract}

\section{Introduction}
A major goal of modern astrophysics is determining how mass and energy circulate between stars
and the interstellar medium. This crucial process takes place in the circumstellar environment.
The available evidence indicates that neither the physics nor the geometry of this process is
simple, and magnetic fields are fundamentally important.  Unfortunately, few magnetic
diagnostics are useful in this environment.  Table 1 lists the common magnetic diagnostics,
together with the component of the field they are sensitive to, whether they measure field
strength or only geometry, the wavelengths used and field magnitudes measurable, additional
requirements for their use, and the general area of astrophysics where they have found
application. Two diagnostics, the Zeeman Effect and gyrocyclotron radiation, have been used, but
they are sensitive only to unusually strong fields (greater than 100 Gauss).  As a result, existing
models of the dynamics of circumstellar matter are left with no observational constraints on the
magnetic field.  In this paper we discuss "scattering" diagnostics, which have heretofore been
applied only to the Sun.  The Hanle Effect is potentially sensitive to moderate (0.1 to 300 Gauss)
fields predicted for winds in some hot stars, for instance, and what we call "magnetic
realignment" is sensitive to quite weak fields (1 $\mu$Gauss to 0.1 Gauss) that would be
important in the outer circumstellar environment.
\begin{table}
\caption{Magnetic Diagnostics}
\begin{center}
\begin{tabular}{llllllll}
\tableline
Method & Comp & Strength & $\lambda$'s & Field & Requires & Useful for\\
\tableline
Faraday & Long & $N_{e} B \ell \lambda^2$ & rad - & $<\mu$G & Bkg pol & diffuse \\ [-
0.1cm]
\ Rotation & & & \ FIR & \ $\uparrow$ & \ Source & \ ISM \\

Gyrocyclo- & Long & & rad & 0.1- & Ionized gas & stellar \\[-0.1cm]
\ tron Rad'n & & & & \ 10kG & & \ coronae\\

Synchrotron & Trans & & rad - & $\mu$G &  Relativistic & diffuse  \\[-0.1cm]
\ Polarization & & & \ vis & \ $\uparrow$ & \ electrons & \ ISM\\

IR Dust & Trans & & FIR & $\mu$G $\uparrow$ & heated dust & mol.\\[-0.1cm]
\ Emission & & & & & &  \ clouds\\

Interstellar & Trans & & UV - & $\mu$G $\uparrow$  & Dust; Bkg & ISM,\\[-0.1cm] 
\ Polarization & & & \ NIR & & \ source & \ clouds\\

Zeeman & Long & $g_L\lambda B$  & vis - & $\mu$G - & narrow line & ISM, \\[-0.1cm]
\ Effect & & & \ rad & \ kG & \ atmos & \ stars \\

Hanle & $\perp$ & $B/A$ & UV - & mG - & Atom scat & stars,\\[-0.1cm]
\ Effect & Illum & \ (line) & \ NIR & \ kG & \ nearby src & \ winds\\

Magnetic & $\perp$ & $B/F$ & UV - & $<\mu$G & Atom scat  & circum*\\[-0.1cm]
\ Realignment & Illum & \ (illum) & \ NIR & \ $\uparrow$ & \ nearby src & \ material\\

\tableline
\tableline
\end{tabular}
\end{center}
\end{table}

\section{Scattering Magnetic Diagnostics}
The scattering diagnostics are based on the polarization from atomic fluorescent scattering.  Here
we summarize the physics behind the effect, first in the absence of a magnetic field, then adding
the Hanle Effect, fluorescent alignment, and magnetic realignment.
\subsection{Atomic Scattering}
The theory of polarization from atomic scattering is summarized in Ignace, Nordsieck \&
Cassinelli (1997) and treated in detail in Stenflo (1994).  To start, we assume thermodynamic
equilibrium and the absence of a magnetic field (treated below).  In this case any one fluorescent
scattering process (lower state to upper state to some possibly different lower state) can  be
treated as the sum of some fraction  $E_1$  of polarized dipole scattering and the fraction  $(1 -
E_1)$  of unpolarized isotropic scattering.  For illumination by an unpolarized beam, the degree
of linear polarization  $p$  of the scattered beam (our observable) as a function of scattering
angle  $\theta$  is
\begin{equation}
p(\theta) =\threequarters E_1 \sin^2 \theta / (1 - \onequarter E_1 + \threequarters E_1 \cos^2
\theta)
\end{equation}
This "dipole-like" scattering has a polarization with a maximum at $90\deg$ scattering which is
proportional to $E_1$.  For an "unaligned" atom (all ground magnetic substates equally
populated) the "polarizability"  $E_1$  may easily be calculated from formulae in Stenflo which
depend only on the angular momentum  $J$  of the initial level and the  $\Delta J(up)$  and 
$\Delta J(down)$  of the scattering process.  $E_1$  is in the range  $-1 < E_1 < 1$, where
positive denotes polarization perpendicular to the plane of scattering, and negative polarization is
parallel to the plane.

The strongest scattering lines in astronomy, and the easiest to understand spectroscopically, are
the "alkali-like doublet" resonance lines, which arise from the ground state of ions with a single
electron above a closed shell (NaI, CaII, NV, etc).  In this paper we will concentrate on these
lines, giving specific examples for sodium.  For the alkali-like doublets, it turns out that the red
line is always unpolarized, $E_1 = 0$, while the blue line is positively polarized.  This is then a
unique signature for line scattering: an unpolarized red line indicates that there are no other
processes, like electron scattering of line emission, contaminating the line scattering polarization. 
For the blue line, in those atoms without hyperfine structure (OVI, CIV, SiIV, MgII, and CaII),
$E_1 = 0.5$, a very high polarizability, producing a strong polarimetric signal. In any case, the
basic polarization properties of an isolated atom are summarized by a single parameter,  $E_1$,
which is a value that can be calculated using known atomic physics.

All this is of interest to this conference because  the observed polarization law  $p(\theta)$  can
be modified by its environment, namely by the magnetic field, and/ or by the presence of a strong
illuminating flux.  Basically, we can in principle observe  $p(\theta)$  and deduce the
environment of the atom.  There are three potential complications in the analysis.  The first is
optical depth: This has been a serious problem in applications of scattering diagnostics to the
solar atmosphere, so that detailed radiative transfer models and consideration of complex
polarization effects in line wings has been important.  However, this is not the case for many
circumstellar applications.  We shall for now assume $\tau \ll 1$, so that the observed
polarization is just that of the Doppler core from isolated atoms.   A second complication is a
non-point illuminator:  This leads to many different scattering angles from the same atom, which
reduces the polarization in a model-dependent way.  Again, while this is a problem in the solar
photosphere, it is not a serious problem in the optically thin outer parts of most stellar winds. 
Third, the most serious complication is that we measure only the polarization  $p$  and we need
an independent estimate of the scattering angle $\theta$.  In the best case we can observe the
polarization at a known scattering angle (e.g. comets).  More usually, the scattering angle must be
inferred from other information.  The method that we advocate here is the use of velocity
information through observation of the polarimetric line profile.  For instance, for a resolved
purely expanding wind (eg, a planetary nebula), along a particular line of sight the Doppler-
shifted wavelength of the line is uniquely related to the position along the line of sight, and thus
to the scattering angle.  For an unresolved stellar wind, the different lines of sight may be
dissected using a model of the radial dependence of the wind (see Cassinelli, Nordsieck \&
Ignace article in this volume).
\subsection{Hanle Effect}
The Hanle Effect occurs for fields strong enough that the Larmor frequency  $\omega_L$  is of
the same order or greater than the inverse lifetime  $A_{ul}$  of the upper level of the atom. 
Classically, the Hanle effect can be treated as a dipole radiator where precession occurs while the
atom is re-emitting the line radiation.  The effect enhances or diminishes the line polarization,
and can lead to a rotation of the angle of the polarization relative to the angle in the case of no
magnetic field. The Hanle effect is significant for fields greater than the "Hanle field" $B_H$,
\begin{equation}
B_H = A_{ul} \,  e \, g_L / m_e \, c  = 5  \, Gauss \times (A_{ul}/10^8),
\end{equation}
where  $g_L$  is the Lande factor and  $A_{ul}$  is the A-value for the upper level.  The Hanle
Effect actually results in a polarization vs scattering angle law that is no longer "dipole-like", and
it now depends not only on  the polarizability $E_1$  but also on  $B_H$  and on the direction
between the magnetic field and the illuminating radiation.  It applies to any flourescent transition
for which $E_1 \neq 0$.  The polarization law varies measurably with field strength when  $B_H
/ 30 < B < 30 B_H$.  It becomes "saturated" (gives angle information only) for larger fields. 
Unlike the Zeeman Effect, the Hanle Effect affects the integrated line, so thermal and bulk
broadening do not destroy it.  This makes it ideal in a dynamic circumstellar environment. 

\begin{figure}
\plotfiddle{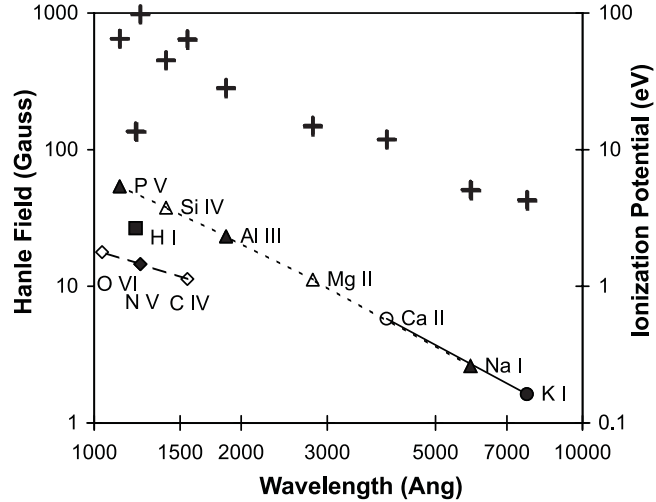}{2.5in}{0}{65}{65}{-200}{-250}
\caption{Left scale.  Hanle Field for the alkali-like doublets, vs wavelength of the line: square:
hydrogen-like; diamonds: lithium-like; triangles: sodium-like; circle: potassium-like. Filled:
alignable due to hyperfine structure.  Right scale (crosses): Ionization potential in eV}
\end{figure}

Figure 1 shows the Hanle Field for all the alkali-like doublets through calcium for atoms with log
abundance greater than 4.5 (H = 12), and for ionization states less than VI.  $B_H$  rises from 3
G for lines in the visible, to 60 G in the far ultraviolet.  These are quite small fields compared to
those detected by the Zeeman Effect, and they are very interesting fields for inner stellar winds. 
The Hanle Effect has been applied in the visible to the Sun, for instance on the sodium D lines. 
For stars hotter than the Sun, sodium is ionized and ions of higher ionization potential (right-
hand scale in figure 1) with lines in the UV and FUV, are more appropriate.  These lines are quite
strong in hot stars and are in fact among those responsible for driving their winds.
\subsection{Hyperfine Structure}
The open symbols (CIV, OVI, MgII, SiIV, and CaII) in Figure 1  are simple doublets; the ground
state is  $^{2}S_{1/2}$  and the upper states are $^{2}P_{1/2}$  and $^{2}P_{3/2}$.  The line
of longer wavelength ("$L_1$") always has $E_1 = 0$, so is unpolarized, and the $L_2$ line
always has $E_1 = 0.5$.  In an interesting complication, the filled symbols in Figure 1, (NV, NaI,
AlIII, PV, and KI), represent ions with nonzero nuclear spin, therefore possessing hyperfine
structure.  For these, the (nondegenerate) hyperfine lines must be treated individually, using total
angular momentum  $F$  instead of  $J$.  It can be shown that the net polarizability for the
$L_1$ line (now a hyperfine multiplet) is still zero, while for $L_2$ it is some nonzero number
that depends on the nuclear spin and on which hyperfine levels are nondegenerate.  For example,
for  $^{23}Na$  (the common isotope of sodium) the nuclear spin is $\frac{3}{2}$, so that the
$^{2} S_{1/2}$ ground state is split into two sublevels, $F = 1,2$; the  $^{2}P_{1/2}$  upper
level is likewise split into two, $F = 1,2$, and the $^{2}P_{3/2}$ upper level is split into four,
$F = 0,1,2,3$.  The sodium nuclear moment is rather large, so these hyperfine levels are all
nondegenerate.  The lower state then has 8 magnetic sublevels (3 for $F=1$; 5 for $F=2$), and
the upper states have 8 and 16, respectively.  The polarizability is calculated (Stenflo 1994) by
summing the contributions for all scattering events beginning and ending in the same magnetic
substates, squaring the result, and then summing these over all allowed initial and final substates. 
For the $D_1$ line one finds  $E_1 = 0$, and for the $D_2$ line, $E_1 = 0.135$, which agrees
with Brossel, Kastler and Winter (1952).

The presence of hyperfine structure has two effects on the alkali doublets.  It does reduce the
polarizability of the $L_2$ line, and thus the magnitude of the Hanle Effect.  But more important,
it makes the atom susceptible to alignment effects which are sensitive to magnetic field in
another way:
\subsection{Fluorescent Alignment}
Any atomic level with  at least three fine or hyperfine levels ($J$ or $F \geq 1$) may be
"aligned".  In the frame work of quantum mechanics, this means that the $M$ substates are not
uniformly populated, and classically it means that in an ensemble of atoms the angular momenta
are not isotropically distributed.  A familiar form of atomic alignment is "optical pumping"
through fluorescence.  This can be understood on the basis of quantum mechanical selection rules
between the $M$  substates of the lower and upper levels.  It will persist as long as the photon
excitation rate  $R_F$  is greater than the collision rate $R_C$ , a situation which holds for many
dilute circumstellar envelopes near bright stars.  For the alkali-like doublets, 
{\it this occurs only for those atoms with hyperfine structure}, where $F \geq 1$.  Atomic
fluorescent alignment has been studied extensively in the laboratory (see Happer, 1972). 
Laboratory studies of sodium  alignment,
pioneered by Brossel, Kastler and Winter (1952) and Hawkings (1954), provide a good
benchmark.

\begin{figure}
\plotfiddle{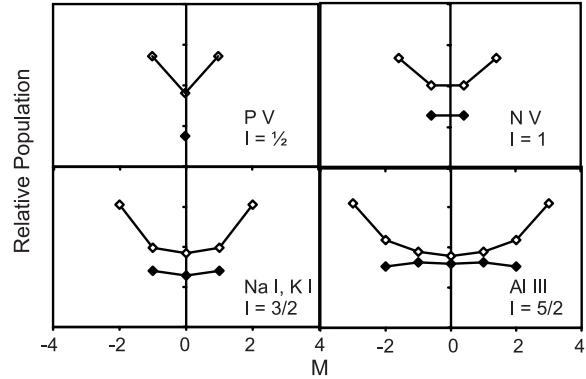}{1.75in}{0}{45}{45}{-150}{-85}
\caption{Relative M state ground level populations for fluorescently aligned alkali-like ions. 
Filled, lower hyperfine level; Open, upper hyperfine level.}
\end{figure}

The importance of fluorescent alignment to this paper is that it (and its modification by a
magnetic field), produces an obervable change in the atomic polarizability.  For the alignable
alkali doublets, the magnitude of the effect depends on the amount of pumping and on the level
structure, which depends on the nuclear spin.  The affect of fluorescent alignment may be
evaluated by a standard statistical equilibrium calculation of the magnetic sublevel population
distribution, where the only transitions are fluorescent transitions illuminated from a single
direction.   Figure 2 shows the population of the $M$ states for the alignable alkali doublets for
complete fluorescent alignment ($R_F \gg R_C$), for nuclear spins $I = \frac{1}{2}$ (PV), $1$
(NV), $\frac{3}{2}$ (NaI, KI) and $\frac{5}{2}$ (AlIII) .  The upper hyperfine level (highest
$F$) is overpopulated, with the states of highest absolute value angular momentum most
overpopulated.  For example, for NaI, $\mid M_F\mid \, = 2$ is overpopulated by 60\%, and the
polarizability of the $D_2$ line goes from 0.135 to 0.217.  Thus aligned Na atoms give a
substantially higher polarization than unaligned ones, by a factor of 1.6, which is easily observed. 

\begin{figure}
\plotfiddle{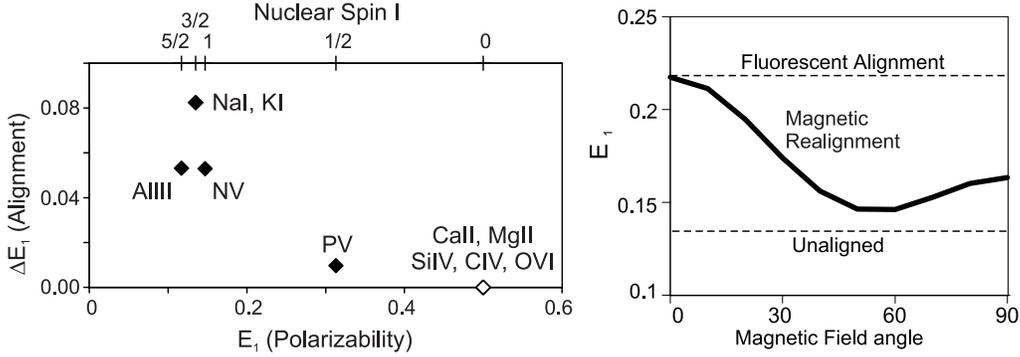}{1.5in}{0}{55}{55}{-215}{-150}
\caption{3a (left), change in polarizability due to fluorescent alignment, vs unaligned
polarizability for $L_2$ line of alkali-like doublets.  3b (right), polarizability as a function of angle
between illumination vector and magnetic field for saturated magnetic realignment of NaD}
\end{figure}

Figure 3a shows a summary for all the alkali doublets.  The vertical axis shows the {\it change in
} polarizability for completely aligned atoms, while the horizontal axis shows the polarizability
for unaligned atoms.
\subsection{Magnetic Realignment}
Fluorescent alignment is interesting to us because it can be altered by a magnetic field.  The
magnetic field mixes the  $M$  sublevels of the atoms, but 
{\it only perpendicular to the field}.   Thus if the Larmor precession rate  $\omega_L$  exceeds
the rate of photon arrival  $R_F$, the atoms are "realigned" to the magnetic field.  This effect we
shall then term "magnetic realignment".  This is once more observable through a change in the
polarizability   $E_1$, which now depends on ratio of the Larmor frequency to the photon arrival
rate, and on the angle of the magnetic field to the photon illumination direction. Recently, the
importance of atomic fluorescent alignment and magnetic realignment for optical transitions has
been realized by solar researchers (see Trujillo Bueno \& Landi Degl'Innocenti 1997). 

In the context of circumstellar envelopes, magnetic realignment is a potentially powerful
magnetic diagnostic, and is analogous to the Hanle Effect in many ways (it is sometimes called
the "second Hanle Effect" in solar physics).  It depends on field strength when $R_F / 30 <
\omega_L < 30 R_F$.  It becomes "saturated" (gives angle information only) for larger fields. 
And it affects the integrated line, so broadening does not destroy it.  Unlike the Hanle Effect,  
{\it it is in principle sensitive to very small fields}.  Defining a characteristic "realignment field"
$B_A$, the field for which $\omega_L \sim  R_F$, we find  
\begin{equation}
B_A  =  (\alpha F_{\nu} / h\nu) (e \, g_L / m_e \, c)  = 0.3  \, \mu Gauss \times (L_{\nu} /L_{\nu}
(sun)) / (R_{AU})^2 ,
\end{equation}
where $\alpha$ is the line cross-section, $F_{\nu}$ and $L_{\nu}$ is the illuminating flux and
luminosity at the frequency of the line, and $R_{AU}$ is the distance in AU from the star. The
realignment field  decreases rapidly in the outer  circumstellar environment.

As an example, we have calculated the effect of saturated magnetic realignment on NaI in the
limit of very small fields ($B_A \ll B \ll B_H$).   For a field oriented at angle  $\phi$  to the
illuminating beam, the effective  $M$  state population may  be calculated using a procedure
given in Hawkins (1954):  The effective  $M$  state population of the ensemble is evaluated by
taking a time average of the frequency of occupation of each  $M$  state over the period between
the arrival of one photon and the next.  Given the new $M$ state population, $E_1 (\phi)$  for
the Na D lines is evaluated.  Once more, for $D_1$ the polarizability remains 0, while for $D_2$
it is intermediate between the unaligned and aligned cases. Figure 3b shows the results for the net
polarization of the $D_2$ line as a function of magnetic field angle.  The observed polarization is
insensitive to magnetic field along the illumination axis, but is quite sensitive for nonzero angles
up to 45\deg .

It is also of interest to evaluate the polarizability of the individual hyperfine transitions within the
D lines.  $D_1$ and $D_2$ are both split into sets of hyperfine lines (termed the "s-resolved"
hyperfine structure), the splitting being about 1 km/s.  This is resolvable in special circumstances,
such as for comets.  In this case all four resolvable lines are polarized for aligned atoms.  The s-
resolved pair for $D_1$ have opposite polarizations which exactly cancel for the net $D_1$ line. 
Measuring the polarization of the s-resolved hyperfine doublets is a potentially powerful
verification of the presence of alignment.
\subsection{A Diagnostic Diagram}
The ultimate result for the scattering diagnostics is that  $p(\theta)$  for an ion depends on the
magnetic field through the affect of realignment on  $E_1 $, and/or through the Hanle Effect
modification of the form of  $p(\theta)$.  In the circumstellar context, the Hanle Effect is
appropriate for moderate fields in usually unresolved inner winds, while magnetic realignment
becomes interesting in outer nebulae (possibly resolved) where it is sensitive to very small fields. 
The Zeeman Effect, on the other hand, is useful in the large fields in or near stellar photospheres.

\begin{figure}
\plotfiddle{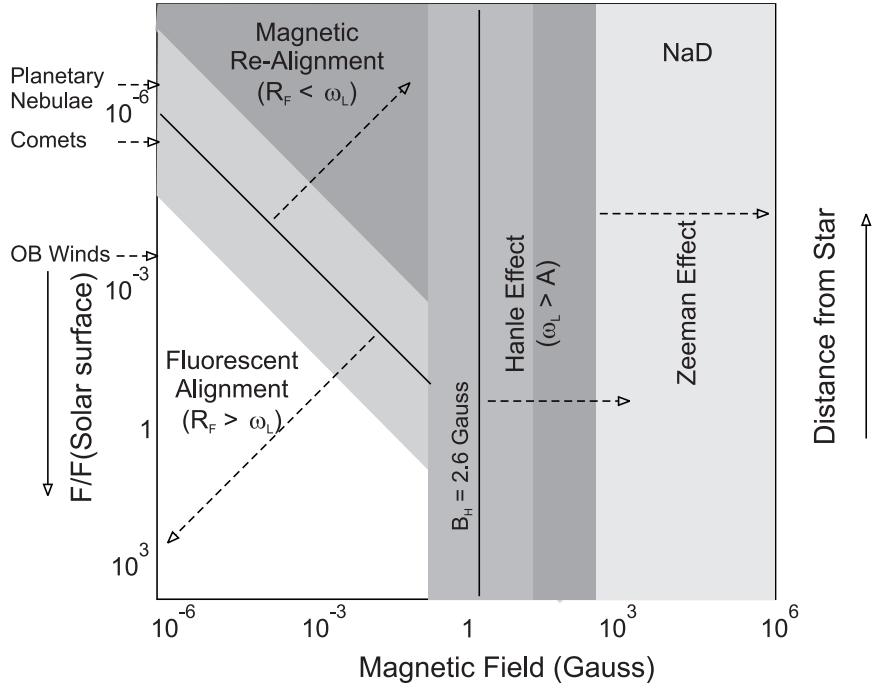}{3.75in}{0}{60}{60}{-175}{-150}
\caption{Diagnostic diagram for circumstellar magnetic fields for Na D.  Vertical axis:
illuminating flux, or distance form illuminator.  Light grey: unsaturated diagnostics, where both
field magnitude and angle may be recovered.  Dark grey: saturated diagnostics, where the
geometry, plus a lower limit to the field magnitude may be recovered.}
\end{figure}

Figure 4 summarizes these magnetic diagnostics.  It applies specifically to NaD, but should be
qualitatively correct for the other hyperfine split alkali doublets (KI, NV).  The vertical axis is the
illuminating flux, normalized to the flux at the surface of the Sun.  In a particular envelope, this
depends just on the distance from the star. Approximate flux levels for the pilot observations
proposed below are shown.  For the spinless alkali-like ions (CaII, CIV, SiIV), only the Hanle
Effect will apply, so that the vertical axis and all the alignment-dependent lines would not
appear, although the Hanle Effect regimes would still appear.  Both magnetic realignment and the
Hanle Effect have regimes (light gray) where both the strength and angle of the magnetic field
may be inferred, and "saturated" regimes (dark gray) where the angle of the field may be deduced
and a lower limit may be placed on the strength.

\section{Instruments/ Investigations}
The scattering diagnostics are still young, so it is not surprising that they have not been applied
beyond the Sun.  Two further barriers are lack of instrumentation and a poor understanding of
how to apply them to unresolved objects.

For instrumentation, a difficulty is the limited spectral resolution of linear polarization
spectropolarimeters. These diagnostics benefit from $R = \lambda /\Delta\lambda > 5000$,
required to resolve lines and to avoid unpolarized continuum contamination and noise.  Second,
because of signal/ noise requirements, one needs larger telescopes and higher efficiency than are
usually available.  Third, some of the diagnostics are better applied in the vacuum ultraviolet,
where polarimetry is still in its infancy.  And finally, for the realignment diagnostic, observations
of faint diffuse lines require unusually high etendue, that is, high spectral resolution on diffuse
targets. 
Below we present two pilot observations together with instrumentation which is being developed
to perform them.
\subsection{Ultraviolet Spectropolarimetry}
Application of the Hanle Effect to hot star winds is an obvious direction to pursue, since the
scattering lines are so prominent, the stars so bright, and the magnetic fields which could be
detected only in this way would be very important in the dynamics of the wind. At the University
of Wisconsin we have been developing the Far Ultraviolet SpectroPolarimeter (FUSP) to explore
this type of observation, among others.   A simulated observation of the O star $\zeta$ Ori with
FUSP is discussed in Cassinelli, Nordsieck, and Ignace in this volume, illustrating the
detectability of a 3 Gauss field with this instrument.

\begin{figure}
\plotfiddle{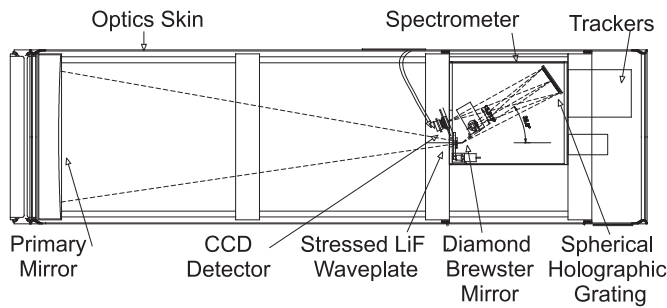}{1.5in}{0}{35}{35}{-150}{-60}
\caption{The Far Ultraviolet SpectroPolarimeter (FUSP)}
\end{figure}
FUSP (figure 5) is a sounding rocket payload designed to obtain the first spectropolarimetry in
the far ultraviolet (Nordsieck, 1999).  It will cover wavelengths 105 - 150 nm with a resolution
$R = 1800$ (0.5 \AA ; 180 km/sec).  The telescope aperture is 50cm and the spectropolarimeter
is at the prime focus.  Polarimetric modulation is provided by a rotating halfwave plate of
stressed lithium fluoride.  The polarization analyzer is a 12 mm square artificial diamond
brewster angle mirror.  The spectrometer uses a spherical aberration-corrected holographic
grating.  A two-stage rocket will carry the payload to an apogee of 400 km, giving a total usable
science time of 400 sec.  The scheduled first launch is in 2002, with a "Hanle Effect" launch
targeting $\zeta$ Ori and the rapid rotator $\xi$ Per in 2003.  FUSP is intended to provide proof
of principle for later development of a small satellite for FUV spectropolarimetry.

\subsection{High Resolution Spectropolarimetric Imaging}
Magnetic realignment appears to be most interesting for the outer parts of circumstellar
envelopes, where its sensitivity to very small fields provides a unique capability.  We are
pursuing as pilot projects observations of resolved nebulae where the scattering angle is most
easily deduced.  There are a number of targets with known NaD scattering nebulae, including
comets, M supergiant stars, and Planetary Nebulae,  which are then observable in the visible,
albeit with large telescopes due to the faintness of the line emission.

\begin{figure}
\plotfiddle{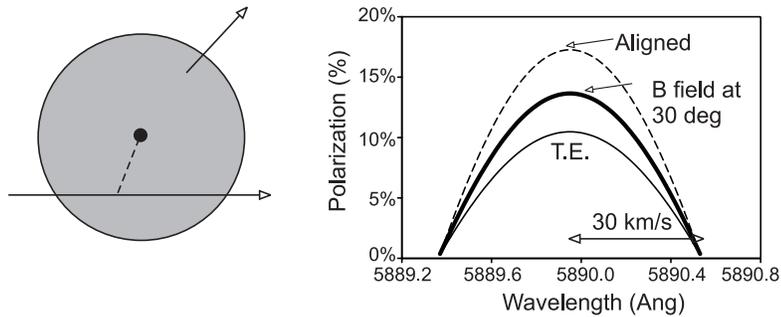}{1.5in}{0}{40}{40}{-170}{-70}
\caption{Simulated polarization profile for a Na $D_2$ line of sight through a Planetary Nebula. 
Left: cartoon showing relation between scattering angle and Doppler velocity.  Right:
polarimetric profile under various assumptions for the magnetic field.}
\end{figure}

An exciting prospect is the detection of magnetic fields in Planetary Nebulae.  The presence of a
field has been suggested by dynamic models of the bipolar geometry of the nebulae.  Field
geometries have been suggested based on models of the evolution of the field in AGB stars
(Thomas, et al, 2001, this volume).   Fluorescent NaD has been seen so far in 5 Planetary
Nebulae by Dinerstein, Sneden \& Uglum  (1995).  The sodium is apparently in the neutral shell
around the nebulae.  At this distance from the central illuminator, the realignment field is less
than 1 $\mu$G, so that any dynamically important field should be easily detected.  Interpretation
of the results is made relatively straightforward by the simple 30 km/s spherical expansion of the
nebula (figure 6). With a resolution of $R > 10,000$ one can resolve the expansion profile. Each
point along the line of sight then corresponds to a unique scattering angle and a unique Doppler
velocity, so that there is a one-to-one correspondence between the position in the profile and the
scattering angle.  The figure shows the expected signal for unaligned atoms (very unlikely),
aligned with no magnetic field, and a saturated magnetic field 30 degrees from the radius vector
to the star. With a map of the polarimetric profile one can in principle obtain two components of
the field geometry in three dimensions. Observations like this will require a large telescope, since
the NaD emission strength is only of order 50 Rayleighs (1 Rayleigh $= 10^6$
photons/sec/4$\pi$ sr) and the nebulae subtend less than 0.5 arcmin.

Such an instrument will become available soon: The University of Wisconsin is constructing a
high-throughput medium resolution imaging spectropolarimeter for the Southern African Large
Telescope (SALT).   SALT will be a 10m telescope based on the Hobby-Eberly Telescope
(HET), which consists of 91 1m segments in a 10m sphere pointed at a fixed zenith distance. 
Sources are tracked for 1- 2 hr using a movable prime focus platform.  The Prime Focus Imaging Spectropolarimeter will operate with a simultaneous
dual visible / NIR beam over 320nm - 1.7\micron;  the visible beam will be commissioned first,
in 2004. Slitmask and long-slit spectropolarimetry will be available over an 8 arcmin field of
view.  The polarimetric modulator will be a superachromatic waveplate ($\onehalf$ and
$\onequarter$ wave), and the analyzer will be a calcite beamsplitter. 
Spectropolarimetry will be available in two modes, grating and Fabry-Perot.  The
grating spectropolarimeter uses Volume Phase Holographic ("VPH") gratings for very high
efficiency.  Resolution will be $R = 800 - 6500$ with an
0.9 arcsec slit (median seeing) and 12,000 with a 0.5 arcsec slit.  Dual etalon Fabry-Perot
spectropolarimetry will also be available with resolutions of 2500 and 13,000.  This seems
most appropriate for the Na D Planetary Nebula pilot project described above.  We estimate that
a single 1 hour observation will obtain a Na D polarimetric "data-cube" with 1\% polarimetric
error/ resolution element over the entire nebula.

\section{Summary}
We find that for the circumstellar environment, the scattering magnetic field diagnostics may be
the most appropriate.  The Hanle Effect used on FUV resonance lines may be valuable for
determining the existence of dynamically important fields (0.1-300 G) in OB star winds. 
Magnetic realignment is a promising magnetic diagnostic of very small fields ($\sim 1 \mu G$)
in the outer circumstellar environment.  However, neither method has been tested due to lack of
appropriate instrumentation.  This is very much work in progress.

\acknowledgements
FUSP is supported by NASA grant NAG5-5252.

\newpage

\section*{Discussion}

\noindent {\bf E. Landi Degl'Innocenti:} Could you please explain in more detail what is the
difference between the Hanle Effect and the phenomenon that you call "realignment?

(related question:)

\noindent {\bf Jan Stenflo:} There is some confusion about the terminology used.  As I
understand it, when you talk about magnetic realignment, what you mean is the magnetic
modification (Hanle Effect) of the ground-state atomic polarization.

\noindent {\bf K. H. Nordsieck:} We have been struggling, as newcomers to the scattering
diagnostics, to apply the experience of solar astronomers to the rather different circumstellar
environment.  I think some of the difficulty is terminology, but that there may be more interesting
insights from the different regimes.  (The following is distilled from discussions later at the
conference and from later thought):  I understand that solar astronomers use the "Hanle Effect" as
a blanket term to cover all effects that involve modification of the polarization through the
"realignment" of the M-states of both the upper and lower levels of a transition.  Solar
astronomers deal with relatively large magnetic fields that are comparable with the Hanle field,
so that there is significant realignment taking place during the scattering process.  I would agree
that in this case it is the same effect for both upper and lower states.  Perhaps we could agree to
call this the "classical" Hanle Effect.  We, on the other hand,  have been exploring the magnetic
realignment of the lower state in a very different regime where the field is much less than the
Hanle Field.  This seems to me to be qualitatively different.  Here, all of the precession of $M$
states occurs {\it before} the scattering process, and because of this, quantum mechanics requires
us to sum results incoherently over an ensemble of atoms in different $M$ states.  I think this
difference explains why in the classical Hanle Effect (field comparable to the Hanle Field), the
polarization phase function becomes non-dipole, with position angle rotations, etc, while in what
I call magnetic realignment (very small field), the polarization remains dipole, with a change only
to the polarizability.

\noindent {\bf G.A. Wade:} Your illustrative calculations regarding the Hanle Effect in $\zeta$
Ori were very interesting.  You took the envelope geometry to be spherical, and hence the
emission is unpolarized in the absence of a magnetic field.  In fact, $\zeta$ Ori, like most other
stars, is rotating.  Its envelope will therefore be somewhat oblate.  How does this effect the
calculations?

\noindent {\bf K.H. Nordsieck:}  Yes, the model is very over-simplified.  One other obvious
problem is that a dipole field in a spherically expanding envelope is not dynamically consistent. 
We picked $\zeta$ Ori as a first example partly because it is not a fast rotator.  It takes quite a lot
of rotation before the wind becomes oblate enough to exhibit intrinsic polarization, so we think
ignoring this is not a serious problem.  That said, if there is sufficient flattening to cause
polarization, it will occur for all scattering lines regardless of their Hanle Field, which should be
separable from the Hanle signal, which is quite sensitive to Hanle Field.  We do plan to look at a
more rapidly rotating O star, $\xi$ Per, to practice dealing with this complication.

\end{document}